\documentclass[prd,twocolumn,aps,psfig,nofootinbib,nobibnotes,superscriptaddress,preprintnumbers,times]{revtex4}
\usepackage{graphicx,bm,color,braket}
\graphicspath{{./fig/}{./png/}}

\newcommand{\BoldVec}[1]{\mathchoice%
  {\mbox{\boldmath $\displaystyle     #1$}}%
  {\mbox{\boldmath $\textstyle        #1$}}%
  {\mbox{\boldmath $\scriptstyle      #1$}}%
  {\mbox{\boldmath $\scriptscriptstyle#1$}}%
}
\newcommand{\EQ}{\begin{equation}}
\newcommand{\EN}{\end{equation}}
\newcommand{\EQA}{\begin{eqnarray}}
\newcommand{\ENA}{\end{eqnarray}}

\newcommand{\Eq}[1]{Eq.~(\ref{#1})}

\newcommand{\Sec}[1]{Sec.~\ref{#1}}

\newcommand{\Fig}[1]{Fig.~\ref{#1}}

\newcommand{\Tab}[1]{Table~\ref{#1}}

\newcommand{\Tabs}[2]{Tables~\ref{#1} and \ref{#2}}

{}
{}
{}

\newcommand{\xx}{\BoldVec{x}{}}

\newcommand{\uu}{\BoldVec{u} {}}

\newcommand{\BB}{\BoldVec{B} {}}

\newcommand{\JJ}{\BoldVec{J} {}}

\newcommand{\ee}{\BoldVec{e} {}}
\newcommand{\ff}{\BoldVec{f} {}}

\newcommand{\kk}{\BoldVec{k} {}}

\newcommand{\nab}{\BoldVec{\nabla} {}}

\newcommand{\SSSS}{\bm{\mathsf{S}}}

\newcommand{\FFF}{\mbox{\boldmath ${\cal F}$} {}}

\def\Rey{\mbox{\rm Re}}

\def\EEK{{\cal E}_{\rm K}}
\def\EEM{{\cal E}_{\rm M}}

\def\EEGW{{\cal E}_{\rm GW}}

\def\OmGW{{\Omega}_{\rm GW}}
\def\hrms{{h}_{\rm rms}}

\def\EErad{{\cal E}_{\rm rad}}

\def\EEcrit{{\cal E}_{\rm crit}}

\def\hc{h_{\rm c}}

\def\xiM{\xi_{\rm M}}

\def\kf{k_{\rm f}}
\def\vA{v_{\rm A}}
\def\urms{u_{\rm rms}}

\def\half{{\textstyle{1\over2}}}

\def\onethird{{\textstyle{1\over3}}}

\newcommand{\MeV}{\,{\rm MeV}}

\newcommand{\uG}{\,\mu{\rm G}}

\newcommand{\Hz}{\,{\rm Hz}}

\newcommand{\nHz}{\,{\rm nHz}}
\newcommand{\uHz}{\,\mu{\rm Hz}}

\newcommand{\K}{\,{\rm K}}

\newcommand{\s}{\,{\rm s}}

\newcommand{\arXiv}[3]{, ``#3,'' arXiv:#2 (#1).}

\newcommand{\yjfm}[4]{, ``#4,'' J. Fluid Mech. {\bf #2}, #3 (#1).}

\newcommand{\yprl}[4]{, ``#4,'' Phys.\ Rev.\ Lett.\ {\bf #2}, #3 (#1).}

\newcommand{\yptrsa}[4]{, ``#4,'' Phil. Trans. Roy. Soc. Lond. A, {\bf #2}, #3 (#1).}

\newcommand{\yapj}[4]{, ``#4,'' Astrophys. J. {\bf #2}, #3 (#1).}

\newcommand{\ypf}[4]{, ``#4,'' Phys. Fluids {\bf #2}, #3 (#1).}

\begin{document}

\title{Can we observe the QCD phase transition-generated gravitational waves through pulsar timing arrays?}

\date{\today}
\preprint{NORDITA-2021-016}

\author{Axel~Brandenburg}
\email{brandenb@nordita.org}
\affiliation{Nordita, KTH Royal Institute of Technology and Stockholm University, 10691 Stockholm, Sweden}
\affiliation{Department of Astronomy, AlbaNova University Center, Stockholm University, 10691 Stockholm, Sweden} 
\affiliation{Faculty of Natural Sciences and Medicine, Ilia State University, 0194 Tbilisi, Georgia}
\affiliation{McWilliams Center for Cosmology and Department of Physics, Carnegie Mellon University, Pittsburgh, PA 15213, USA}

\author{Emma Clarke \footnote{Corresponding author; the authors are listed alphabetically.}}
\email{emmaclar@andrew.cmu.edu}
\affiliation{McWilliams Center for Cosmology and Department of Physics, Carnegie Mellon University, Pittsburgh, PA 15213, USA}

\author{Yutong He}
\email{yutong.he@su.se}
\affiliation{Nordita, KTH Royal Institute of Technology and Stockholm University, 10691 Stockholm, Sweden}
\affiliation{Department of Astronomy, AlbaNova University Center, Stockholm University, 10691 Stockholm, Sweden}

\author{Tina~Kahniashvili}
\email{tinatin@andrew.cmu.edu}
\affiliation{McWilliams Center for Cosmology and Department of Physics, Carnegie Mellon University, Pittsburgh, PA 15213, USA}
\affiliation{Faculty of Natural Sciences and Medicine, Ilia State University, 0194 Tbilisi, Georgia}
\affiliation{Abastumani Astrophysical Observatory, Tbilisi, GE-0179, Georgia}
\affiliation{Department of Physics, Laurentian University, Sudbury, ON P3E 2C, Canada}

\begin{abstract}
We perform numerical simulations of gravitational waves (GWs) 
induced by hydrodynamic and hydromagnetic
turbulent sources that might have been present at cosmological 
quantum chromodynamic (QCD) phase transitions. 
For turbulent energies of about 4\% of the radiation energy density,
the typical scale of such motions may have been a sizable fraction of
the Hubble scale at that time.
The resulting GWs are found to have an energy fraction of about $10^{-9}$
of the critical energy density in the nHz range today and may already
have been observed by the NANOGrav collaboration.
This is further made possible by our findings of shallower spectra 
proportional to the square root of the frequency for nonhelical
hydromagnetic turbulence.
This implies more power at low frequencies than for the steeper spectra
previously anticipated.
The behavior toward higher frequencies depends strongly on the nature of the turbulence.
For vortical hydrodynamic and hydromagnetic turbulence, there is a sharp drop of spectral GW energy
by up to five orders of magnitude in the presence of helicity, and somewhat
less in the absence of helicity.
For acoustic hydrodynamic turbulence, the sharp drop is replaced by a power law decay,
albeit with a rather steep slope.
Our study supports earlier findings of a quadratic scaling of the GW
energy with the magnetic energy of the turbulence and inverse quadratic 
scaling with the peak frequency, which leads to larger GW energies under QCD conditions.
\end{abstract}

\maketitle

\section{Introduction}
Gravitational wave (GWs) astronomy opens a new window to study the physical
processes in the early universe.
Relic GWs can be sourced by violent processes such as 
cosmological phase transitions and after their generation they propagate
almost freely throughout the expansion of the universe that causes
the dilution of their strain amplitude and frequency; see 
Refs.~\cite{Hogan:1986qda,Krauss:1991qu,Signore:1989,
Kosowsky:1992rz,Kamionkowski:1993fg} for pioneering works and
Ref.~\cite{Caprini:2018mtu} for a review and references therein.
On the other hand, the detection of these relic GWs
is a challenging task due to their small amplitudes, the specific
range of the characteristic frequencies, and astrophysical foregrounds
\cite{Romano:2016dpx}. 
Despite tremendous advancements in  
GW detection techniques, the stochastic GW background 
of cosmological origin remained unobserved.

Recently, the NANOGrav collaboration reported strong evidence for a stochastic 
GW background \cite{Arzoumanian:2020vkk}. In addition to the possibility of GWs induced by astrophysical sources
such as supermassive black holes, the NANOGrav data can also be 
understood
as a possible signal from the early universe, 
such as inflationary GWs
\cite{Kuroyanagi:2020sfw,Tahara:2020fmn,
Cai:2020qpu,Zhou:2020kkf,Vagnozzi:2020gtf,Sakharov:2021dim},
cosmic strings and domain walls
\cite{Ellis:2020ena,Buchmuller:2020lbh,Samanta:2020cdk,
Liu:2020mru,Paul:2020wbz,Blasi:2020mfx},\footnote{
Some tension between NANOGrav limits and Parkes PTA (PPTA) 
has been discussed in Ref.~\cite{Lazarides:2021uxv}}.
inhomogeneous neutrino plasma \cite{Pandey:2020gjy,Pandey:2019tmo},
phase transitions including the supercooled
phase transitions \cite{Lewicki:2020azd}, dark phase
transitions 
\cite{Nakai:2020oit,Addazi:2020zcj} and quantum
chromodynamic (QCD), with axionic string network and QCD axion
\cite{Kitajima:2020rpm,Ramberg:2020oct,Lee:2020wfn,Gorghetto:2021fsn}, 
and/or magnetic fields
\cite{Neronov:2020qrl} and turbulence \cite{Abe:2020sqb}. 
In fact, the observed 45 pulsars from the NANOGrav 12.5-yr dataset
were used to search for cosmological first order phase transitions that
occur below the electroweak energy scale \cite{Arzoumanian:2021teu}
and might require physics beyond the standard model.
However, since there is a degeneracy with the supermassive black hole
signal, distinguishing cosmological sources from the astrophysical ones
is a complex task \cite{Romano:2016dpx,Biscoveanu:2020gds} that can be
accomplished through different observational data cross correlations
\cite{Moore:2021ibq}.
The most promising is astrometric data from current and nearest future
missions \cite{Garcia-Bellido:2021zgu}.
Interestingly, the search of GWs through astrometry includes polarization 
state measurements \cite{OBeirne:2018slh}, and correspondingly to
constrain non-standard models of gravity \cite{Cornish:2017oic}.

In this paper we present a self-consistent study of the 
GWs from turbulent sources possibly present at QCD phase transitions. 
We extend the work of Ref.~\cite{Neronov:2020qrl} by considering
a range of typical length scales of the turbulent motions. Such turbulent 
sources might be driven not just by magnetic fields 
\cite{Quashnock:1988vs,Cheng:1994yr,Sigl:1996dm,Forbes:2000gr,
Kisslinger:2005uy,Urban:2009sw,Buividovich:2009wi,Boyanovsky:2002wa}, 
but could include other turbulent sources at the QCD energy scale such
as the aforementioned combined effect of QCD axions and magnetic fields
\cite{Ramberg:2020oct}.

We also investigate the dependence of spectral amplitude and shape on 
the number of eddies (around 100 at electroweak phase transitions 
and 5-6 at the QCD phase transitions) within a linear Hubble scale.

As we show, this dependence might be crucial when considering the shape
of the GW spectra at low frequencies, as well as the resulting GW energy
density strength.
Even if primordial fields are not dynamically strong, 
turbulence can still develop at QCD energy scales 
\cite{Quashnock:1988vs,Rischke:2003mt,Kiskis:2003rd,Aoki:2006we,
Kahniashvili:2009mf,Blaizot:2013hx,Miniati:2017kah}; the latent 
heat they release still gives rise to pressure gradients resulting in
macroscopic plasma motions.
Given the very high Reynolds number of the primordial plasma 
\cite{Ahonen:1996nq}, such
motions will inevitably decay into turbulence \cite{Quashnock:1988vs,Miniati:2017kah}. 
As already alluded to above, particularly
important for our work is the earlier finding that
the separation and size of nucleation bubbles in a QCD phase
transition is a sizeable fraction of the Hubble scale; see 
Refs.~\cite{Hogan:1983zz,Witten:1984rs} for pioneering works and follow-up papers
\cite{Applegate:1985qt,Midorikawa:1985gy,Strumia:1998nf,
Cottingham:1993rv,Hindmarsh:1991ay,Ignatius:1993qn,Ignatius:1997va,
Strumia:1999fv,Schwarz:2003du,Tawfik:2011sh,Tawfik:2011mw}. 
Furthermore, the assumption of turbulence being driven by magnetic fields,
allows us to avoid the requirement of first order QCD phase transitions
\cite{Miniati:2017kah}. 

The paper is organized as follows.
We first review basic properties of relic GWs (\Sec{relicGWs}),
discuss then the NANOGrav observations (\Sec{NANOGravDATA}),
present our numerical approach (\Sec{ModelingGWs})
and results (\Sec{NumericalSimulations}) of our simulations,
before concluding in \Sec{Concl}.
Throughout the paper, we use natural units with  $\hbar=c=k_B=1$. 
We also set the permeability of free space to unity, i.e., $\mu_0=1$,
expressing the electromagnetic quantities in Lorentz-Heaviside units. 
The Latin indices run $i \in (1,2,3)$ and define the spatial coordinates, and the Greek indices run $\lambda \in (0,1,2,3)$. 
We choose the metric signature as $(-1,1,1,1)$. 

\section{The Early-Universe Gravitational Wave Signal}
\label{relicGWs}

GWs correspond to the tensor mode of perturbations $\delta g_{\mu\nu}$ above the spatially
flat, homogeneous, and isotropic Friedmann-Lema\^{i}tre-Robertson-Walker
(FLRW) background, in the transverse-traceless (TT) gauge\footnote{The 
TT gauge is determined by the TT projection tensor $\Lambda_{ijkl}
= P_{ik}P_{jl} - \frac{1}{2}P_{ij}P_{kl}$, where the $P_{ij}$ is a
transverse operator ($\partial_i P_{ij} =0$), defined as
$P_{ij} = \delta_{ij} - \partial_i \partial j/{\nab}^2$, where 
$\delta_{ij}$ is the Kronecker delta, $\partial_\lambda \equiv \partial/\partial x^\lambda$
denotes the partial derivative in respect of $x^\lambda$ coordinate, 
and $(\nab$ defines the vector differential operator with the components
equal to ${\nab}_i \equiv \partial_i$, i.e., ${ \nab}^2$ is the 
Laplacian in respect of spatial coordinates;
see for more details Chapter 1 (1.2) of \cite{MaggioreText} 
} 
defined through the spatial component $h_{ij}^{\rm phys}$ with 
$a^2 h_{ij}^{\rm phys}= \Lambda_{ijlm} \delta g_{lm}$, where 
$a$ is the scale factor at the physical time $t_{\rm phys}$.
Here and below, super/subscript ``phys" denotes physical quantities.

In order to eliminate the expansion-induced dilution
from the governing hydromagnetic equations, we use rescaled 
quantities together with the conformal time $t$, defined through $dt=dt_{\rm phys}/a$, 
which reduces the metric tensor to the Minkowski form.
The background expansion of the universe during the radiation-dominated 
epoch is governed by the (dominant) radiation energy density
$\EErad = {\pi^2 g(T) T^4}/{30}$, where $g(T)$
is the effective number of relativistic degrees of freedom at temperature $T$. 
In the epoch(s) of interest, the expansion of the universe is fully governed by radiation, and 
the Hubble parameter $H \equiv a^{-1}da/dt_{\rm phys} = a^{-2}da/dt$ 
is given through $H^2 (t) = (8\pi G /3)\, \EErad (t)$,
where $G$ is Newton's gravitational constant and 
$\EErad(t)$ denotes the total energy density
of radiation (including all relativistic components).

In order to connect physical and comoving variables and to determine the
scaling of physical quantities, we compute the ratio of the scale factor 
today, $a_0=a(t_0)$ (here and below, ``0'' denotes the present moment),
to that at the time $t_*$ (at the temperature $T_*$ at which the source
becomes active and the gravitational signal is generated) corresponding
to the start of the simulation. We assume the adiabatic expansion of
the universe, such that $g_S(T)T^3a^3(T)$ is constant, where $g_S(T)$
is the number of adiabatic degrees of freedom at temperature $T$.
At high enough temperatures ($T>1\MeV$), we have $g_S(T)=g(T)$ \cite{K&T}.
Note, that our consideration below is valid for any time period 
during the radiation dominated epoch. However, we will be focused 
on the time period around the QCD energy scale ($150\MeV$). 
We also normalize the scale factor $a_* \equiv a(t_*)=1$,
which differs from the usual convention $a_0 =1$. 
Entropy conservation leads to
\begin{equation}
    \frac{a_0}{a_*} = 10^{12} \, \bigg(\frac{g_{S}(T_*)}{15}\bigg)^{1/3} 
    \bigg(\frac{T_*}{150\MeV}\bigg),
\label{Dfactor}
\end{equation} 
where we have used $T_0 = 2.73\K$ and $g_S(T_0) = 3.91$,
while at the QCD energy scale we have $g_S(T_*) \approx 15$ \cite{K&T}.
The degrees of freedom at QCD is approximate due to uncertainty in the
exact temperature of the QCD transition and knowledge of the standard
model (see discussions in \cite{Schwarz:2003du,Husdal:2016haj}).
However, as $(a_0/a_*)\sim g_S(T_*)^{1/3}$, small deviations in $g_S(T_*)$
will not significantly impact our results.

The GW equation in physical time and space coordinates is given by
\begin{equation}\label{eq:GWeqn1} 
    \big(\partial_{t_{\rm phys}}^2 + 3H\partial_{t_{\rm phys}}- \nab^2_\mathrm{phys}\big)  
    h^\mathrm{phys}_{ij} = 16\pi G  T^{\rm TT}_{ij,{\rm phys}},
\end{equation}
where the TT superscript denotes the TT projection 
of the stress-energy tensor 
such that $T^{\rm TT}_{ij,{\rm phys}} = \Lambda_{ijlm} T^{\rm phys}_{lm}$. 

To make the connection with observations, we define 
the characteristic strain, $h_c(t)$, which
obeys $h_c^2 (t)=\braket{(h_{ij}^{\rm phys}(\mathbf{x},t))^2}/2$, 
where angle brackets denote volume averaging in physical space, 
and the physical energy density $\EEGW^{{\rm phys}}(t)$ 
carried by the GWs is given by \cite{MaggioreText}
\begin{equation}
    \EEGW^{{\rm phys}}(t) = \frac{1}{32\pi G}\braket{(\partial_{t_\mathrm{phys}}h_{ij}^{\rm phys}(\mathbf{x},t))^2}.
\end{equation} 
It is then expressed in terms of today's
frequency $f = k / (2\pi a_0)$  that corresponds to the time Fourier transform
$Q(t)= \int_{-\infty}^\infty df \, Q(f) e^{-2\pi f t}$ 
(and $Q(f)= 2\pi \int_{-\infty}^\infty dt \, Q(t) e^{-2\pi f t}$) 
\cite{MaggioreText}.

The relic GW signal strength today is 
given through the normalized GW energy 
density parameter 
$\OmGW(f)$ reduced by the factor $(H_*/H_0)^2 (a_*/a_0)^4$, 
where $H_*$ is the Hubble parameter at $t_*$. 
This accounts for the dilution of the GW energy density parameter with
the expansion of the universe and renormalizes the GW energy density
by the critical energy density at the present time, $\EEcrit^0 =
(3H_0^2)/(8\pi G)$, where $H_0 = 100 \, h_0$ km s$^{-1}$ Mpc$^{-1}$
$\simeq 3.241 \times 10^{-18} \, h_0 \, \mathrm{s}^{-1}$ is the present
value of the Hubble parameter. 
A frequency of particular interest is 
the frequency $f_*$ corresponding to the Hubble horizon scale at $t_*$:
\begin{equation}
    f_* = \frac{a_*H_*}{a_0} \simeq (1.8 \times10^{-8}\;\mathrm{Hz})\bigg(\frac{g_*}{15}\bigg)^{1/6}\bigg(\frac{T_*}{150\MeV}\bigg).
\end{equation}

As discussed above, there is a variety of possible sources of a stochastic
GW background in the nHz frequency range, accessible to Pulsar Timing
Arrays (PTAs), see Refs.~\cite{Sazhin:1978,Detweiler:1979wn} 
for pioneering works; see Section~\ref{NANOGravDATA} for more details and
Ref.~\cite{Burke-Spolaor:2018bvk} for a review and references therein,
and these sources include a cosmic population of supermassive
black hole binaries (SMBHBs) \cite{Sesana:2008mz,Burke-Spolaor:2018bvk}, 
cosmic strings \cite{Sanidas:2012ee,Cutler:2013aja,Blanco-Pillado:2017rnf},
inflationary GWs \cite{Campeti:2020xwn,Vagnozzi:2020gtf},\footnote{The quantum
mechanical fluctuations during the inflationary epoch induces GWs 
via parametric resonance 
\cite{Grishchuk:1974ny,Rubakov:1982df,Starobinsky:1979ty}.}
any anisotropic stress possibly present in the early universe 
\cite{Deriagin:1987}, and phase transitions in the early universe
(around the QCD energy scale); see, e.g.,
\cite{Signore:1989,Ellis:2012in,Capozziello:2018qjs,Schettler:2010wi,
Kahniashvili:2009mf,Ramberg:2020oct,Caprini:2010xv,Kobakhidze:2017mru,Thorsett:1996dr}.

We also present upper limits on the relic (prior to recombination) 
GW background strength based on Big Bang nucleosynthesis (BBN) 
and the cosmic microwave background (CMB), as well as theoretically
estimated strength and characteristic frequencies for different sources;
see Sec.~\ref{NumericalSimulations}.
The estimated characteristic frequency and wave number of GWs are 
related to each other through $2\pi f=k$,
and can be expressed in terms of the characteristics
(length and time scales) of the source.
In particular, if we assume that GWs could be sourced by bubble collisions at a phase transition, 
we expect the frequency of the GWs to be related to the bubble size. 
We consider that the bubble length scale is the Hubble horizon $H_*$ 
at generation divided by the total number of phase transition bubbles $N_b$. 
Then, for the QCD phase transitions, the frequency is given by
\begin{equation}
    f_* \simeq (1.1 \times10^{-7}\;\mathrm{Hz})\bigg(\frac{g_*}{15}\bigg)^{1/6}\bigg(\frac{T_*}{150\MeV}\bigg)\bigg(\frac{N_b}{6}\bigg),
\end{equation}
where we have normalized to 6 bubbles expected at the QCD phase transition \cite{Schwarz:2003du}. 
This argument applies to the case of a first order QCD phase transition.
Alternatively, it has been proposed to explain the stochastic GW background
from magnetogenesis \cite{Sharma:2018kgs} in low energy scale reheating around the 
QCD epoch
\cite{Sharma:2021rot}.
In addition, through the axion-driven turbulence generation scenario 
\cite{Miniati:2017kah}, a first order phase transition is not required 
and the number of eddies or bubbles ($N_b$) defines the size of the 
largest turbulent eddy that was excited through the axion-driven mechanism. 
We present our result without specifying the number of bubbles, noticing
that more detailed consideration is required to determine the axion-driven 
turbulence characteristics, which is beyond the scope of the current paper.

\section{NANOGrav data}
\label{NANOGravDATA}

A pulsar is a highly magnetized and rapidly rotating neutron star that
emits a beam of electromagnetic radiation along its magnetic axis \cite{Ruderman:1975ju}. 
The times of arrival (TOA) of these pulses are extremely regular and
can be predicted very accurately over long times \cite{Gold:1968}.
The presence of a GW passing between the observer and pulsar shifts
the pulse TOA proportional to the amplitude of the GW 
\cite{MaggioreText2}. 
By monitoring the fluctuations in the TOA of radio pulses from millisecond 
pulsars (see, e.g., for a review Ref.~\cite{Taylor:2021yjx} and reference
therein and for identifying noise sources in PTA see
Ref.~\cite{Goncharov:2020krd} and references therein)
international PTA missions\footnote{The International Pulsar Timing
Array (IPTA) is a consortium of consortia, comprised of the European
Pulsar Timing Array (EPTA), the North American Nanohertz Observatory
for Gravitational Waves (NANOGrav), and the Parkes Pulsar Timing Array
(PPTA) \cite{PTA}.} aim to probe a stochastic GW background.

The maximum sensitivity of a PTA experiment is limited by the total observation time.
That is, the lowest detectable frequency is on the order of the inverse
of the time span of the data (e.g., $f \sim$ nHz for datasets spanning
$\sim$ 10 years) \cite{Hobbs:2009yy}.
Furthermore, data sampling (i.e., pulsars are usually observed on the order
of weeks \cite{MaggioreText2}) limits the maximum detectable frequency. The NANOGrav 12.5-year data is sensitive to GW frequencies between approximately
$2.5\nHz$ and $1 \uHz$ \cite{Brazier:2019mmu}.

PTA measurements typically characterize a stochastic GW background in
terms of its characteristic strain spectrum $h_c(f)$ fitted with a
power-law dependence on frequency
\cite{Arzoumanian:2020vkk},
\begin{equation}\label{eq:hfit}
    \hc(f)=A_\mathrm{CP}\bigg(\frac{f}{f_\mathrm{yr}}\bigg)^{\alpha_\mathrm{CP}},
\end{equation}
where the subscript ``CP" denotes a common-spectrum (CP) process (common to the observed pulsars), the spectral index $\alpha_\mathrm{CP}$
depends on the source of the stochastic GW background, and $A_\mathrm{CP}$
is the strain amplitude at a reference frequency of
$f_\mathrm{yr} = 1 \, \mathrm{yr}^{-1}$.
This choice of reference frequency is arbitrary and does not affect the
ability to detect a GW signal.

The energy density spectrum of the GW background today expressed in terms
of the characteristic strain spectrum is given by \cite{MaggioreText2}
\begin{equation}\label{eq:OmGWhc}
    \OmGW(f)=\frac{2\pi^2}{3H_0^2}f^2 h_c^2(f) = \OmGW^{\rm yr}\bigg(\frac{f}{f_\mathrm{yr}}\bigg)^{5-\gamma_\mathrm{CP}},
\end{equation} 
where we have used Eq.~(\ref{eq:hfit}) in the second term on the right-hand side, 
$\gamma_\mathrm{CP}=3-2\alpha_\mathrm{CP}$ and
$\OmGW^{\rm yr} \equiv 2\pi^2A_{\rm CP}^2f_{\rm yr}^2/(3H_0^2)$.
The quantity $h_0^2\OmGW(f)$ is typically considered in order to remove 
the uncertainty in the value of $H_0$.

The NANOGrav collaboration reports joint $A_\mathrm{CP}-\gamma_\mathrm{CP}$ posterior distributions \cite{Arzoumanian:2020vkk}.
Posteriors for a common-spectrum process in the NANOGrav 12.5-year data were recovered with four models:
free-spectrum, broken power law, 5-frequency power law, and 30-frequency power law. 
The fits were performed for frequencies $f \in [2.5\times10^{-9}, 7\times10^{-8}]\Hz$,
with the exception of the 5-frequency power law, which was fit to the five lowest frequency bins.
The four lowest frequency bins have the strongest response to the presence of a GW background 
(see Figure~13 of Ref.~\cite{Arzoumanian:2020vkk}).
Thus, the 5-frequency power law was fit within
the signal-dominated frequency range (approximately $f \in [2.5\times10^{-9}, 1.2\times10^{-8}]\Hz$).
Figure 1 of 
Ref.~\cite{Arzoumanian:2020vkk} shows the 1$\sigma$ and 2$\sigma$ posterior contours
for the amplitude $A_\mathrm{CP}$ and spectral slope $\gamma_\mathrm{CP}$.

\begin{figure} 
\centering
\includegraphics[scale=0.8]{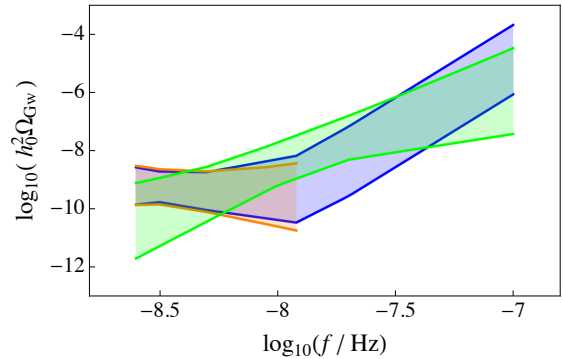}
\caption{NANOGrav 12.5-year data set $2\sigma$ confidence
contours for the posteriors of a common-spectrum process
(see Ref.~\cite{Arzoumanian:2020vkk} Figure 1) expressed in terms of
the GW energy density $h_0^2\Omega_\mathrm{GW}(f)$ and frequency $f$.
This is shown over the NANOGrav 12.5-year sensitivity range of 2.5--100 nHz.
The three models used to fit the process include a: broken power law
(blue), 5-frequency power law (orange), and 30-frequency power law
(green).}
\label{fig:NGcontours}
\end{figure}

Fig.~\ref{fig:NGcontours} shows the NANOGrav detection expressed
in terms of $h_0^2\Omega_\mathrm{GW}(f)$ as given by 
Eq.~(\ref{eq:OmGWhc}).
The shaded regions show the 2$\sigma$ confidence contours of the
$A_\mathrm{CP}-\gamma_\mathrm{CP}$ parameter space in terms of $f$ and
$h_0^2\Omega_\mathrm{GW}(f)$ for frequencies from 2.5--$100\nHz$
(i.e., the NANOGrav 12.5-year sensitivity range);
see Ref.~\cite{Arzoumanian:2020vkk} for more detail.

\section{Gravitational Wave Generation} 
\label{ModelingGWs}

As mentioned in the introduction (see also Ref.~\cite{Burke-Spolaor:2018bvk}),
low frequency GWs can originate from various astrophysical foreground
sources (white dwarfs, SMBH mergers, etc),
and from relic sources related to inflation and cosmic strings, for
example, and in particular, from phase transition-generated turbulence
and primordial magnetic fields.
We focus here on the latter two.
Turbulence and/or magnetic fields would only be generated during a
limited amount of time before they would decay.
The decay process itself remains highly turbulent and could
affect GW production.
Let us therefore begin with some general remarks about turbulent decay. 

\subsection{Gravitational Waves from Turbulent Sources}

Using scaled quantities $h_{ij}=a h^{\rm phys}_{ij}$ and
$T_{ij}^{\rm TT}=a^4 T^{\rm TT}_{ij,{\rm phys}}$, together with 
$a(t) \propto t$
in the radiation dominated epoch, 
the GW equation takes the form
\begin{equation}\label{eq:GWeqn}
    \left(\partial_t^2 - \nabla^2\right)h_{ij} = \frac{16\pi G}{a} T^{\rm TT}_{ij}.
\end{equation}
To obtain the GW equation in Fourier (wavenumber)
space, we use the Fourier transforms and the polarization $r=(+,\times)$ decomposition of the 
tensor metric perturbations and stress energy tensor projected onto the TT gauge  (i.e.,
${\cal Q}_{ij}(\kk,t) = \sum\limits_{r=+,\times} e^r_{ij}(\hat{\kk}) Q_{ij}(\kk,t)$,
where $e^{+}_{ij}(\hat{\kk})$ and
$e^\times_{ij}(\hat{\kk})$ are the polarization tensors with 
$\hat{\kk}$ the unit vector, and $\kk = a \kk_{\rm phys}$ is the 
rescaled 
wavenumber).\footnote{
We use the spatial Fourier transform convention: 
${\cal Q}(\xx,t) = \int \frac{d^3\kk}{(2\pi)^3}e^{-i\kk\cdot\xx}{\cal Q}(\kk,t)$ 
and ${\cal Q}(\kk,t) = 
\int d^3\xx e^{-i\kk\cdot\xx}{\cal Q}(\xx,t)$. 
The transverse operator $P_{ij}$ in the Fourier
space is given $P_{ij}(\hat{\kk}) = \delta_{ij} -
\hat{k}_i\hat{j}_k$ 
and the TT projection operator is 
$\Lambda_{ijkl}(\hat{\kk}) = P_{ik}(\hat{\kk})P_{jl}(\hat{\kk})
-\frac{1}{2}P_{ij}(\hat{\kk})P_{kl}(\hat{\kk})$, correspondingly.
The polarization tensors  $e^{+}_{ij}(\hat{\kk})$ and $e^{\times}_{ij}(\hat{\kk})$ can be written as $e^+_{ij}(\kk) =
{\bf \hat{e}}^1_i{\bf \hat{e}}^1_j - {\bf \hat{e}}^2_i{\bf \hat{e}}^2_j$ and $e^\times_{ij}(\kk)
= {\bf \hat{e}}^1_i{\bf \hat{e}}^2_j + {\bf \hat{e}}^2_i{\bf \hat{e}}^1_j$, where ${\bf \hat{e}}^1$
and ${\bf \hat{e}}^2$ are unit vectors that are 
orthogonal to $\hat{\kk}$  
and each other; see 
Chapter 1 (1.2) of Ref.~\cite{MaggioreText}. 
and Ref.~\cite{Pol:2018pao} for further detail.} 

As in earlier work \cite{Pol:2018pao,Pol:2019yex}, we use normalized
conformal time, $\bar{t}=t/t_*$, where $t_*=H_*^{-1}$ is our starting
time, and $a_*=1$ has been chosen.
Therefore, $a=\bar{t}$.
We also use the scaled wave vector, 
$\bar{\kk}=\kk/H_*$, 
and a scaled 
normalized
stress, $\bar{T}_{+/\times}^{\rm TT} = T_{+/\times}^{\rm TT}/\EErad^*$. 
The GW equation can then be written in the form \cite{Pol:2018pao,Pol:2019yex}
\begin{equation}
\left( \partial^2_{\bar{t}} + \bar{\kk}^2\right)
h_{+/\times} (\kk, t)\, =\, {6 \over \bar{t}}
\bar{T}_{+/\times}^{\rm TT} (\kk, t),
\label{GW}
\end{equation}
but from now on we omit all overbars.

Throughout this paper, all numerical results will usually be presented
as the scaled variables introduced above.
In particular, we quote the rms strain, $\hrms=\braket{h^2}^{1/2}$,
where $h^2=h_+^2+h_\times^2=(h_{ij})^2/2$, and likewise for the
scaled GW energy, $\EEGW=\braket{\dot{h}^2}/6$, 
where $\dot{h}_{+/\times}=\partial_t h_{+/\times}$
with $\dot{h}^2\equiv\dot{h}_+^2+\dot{h}_\times^2$;
see Ref.~\cite{Pol:2018pao,Pol:2019yex} for additional
subdominant terms that are applied in the calculations. 
We sometimes also quote the (frequency dependent) characteristic
amplitude of the physical strain measured today, $h_c(f)=\hrms/a_0$;
see \Sec{relicGWs}.

\subsection{Turbulent Sources}

Turbulent flows in the early universe can be modeled by solving the
hydromagnetic equations for the density $\rho$, the velocity $\uu$,
and the magnetic field $\BB$ with $\nab\cdot\BB=0$, adopting an 
ultrarelativistic equation of state in an expanding universe
using conformal time and comoving variables
\cite{Brandenburg:1996fc,Brandenburg:2017neh} with a forcing
term $\FFF$ in the induction equation for $\BB$
\begin{eqnarray}
{\partial\ln\rho\over\partial t}
&=&-\frac{4}{3}\left(\nab\cdot\uu+\uu\cdot\nab\ln\rho\right)
+{1\over\rho}\left[\uu\cdot(\JJ\times\BB)+\eta \JJ^2\right],
\nonumber \\
{\partial\uu\over\partial t}&=&-\uu\cdot\nab\uu
+{\uu\over3}\left(\nab\cdot\uu+\uu\cdot\nab\ln\rho\right)
+{2\over\rho}\nab\cdot\left(\rho\nu\SSSS\right) \nonumber \\
&&-{1\over4}\nab\ln\rho -{\uu\over\rho}\left[\uu\cdot(\JJ\times\BB)+\eta\JJ^2\right] +{3\over4\rho}\JJ\times\BB,
\nonumber \\
{\partial\BB\over\partial t}&=&\nab\times(\uu\times\BB-\eta\JJ+\FFF),\quad
\JJ=\nab\times\BB.
\nonumber
\end{eqnarray}
We recall that the conformal time $t$ 
is normalized to unity at the time $t_*$
of magnetic field generation, $\rho$ is in units of the initial value,
$\uu$ is in units of the speed of light, and the magnetic energy density
$\BB^2/2$ is measured in units of the radiation density at the time
of generation.
Furthermore, 
${\sf S}_{ij}=\half(u_{i,j}+u_{j,i})-\onethird\delta_{ij}\nab\cdot\uu$
are the components of the rate-of-strain tensor with commas denoting
partial derivatives, $\JJ$ is the current density,
$\nu$ is the kinematic viscosity, and $\eta$ is the
magnetic diffusivity.
The electromotive force, $\FFF$, is used to model magnetic field
generation with
\begin{equation}
\FFF(\xx,t)={\rm Re}[{\cal N}\tilde{\ff}(\kk)\exp(i\kk\cdot\xx+i\varphi)],
\label{emf_model}
\end{equation}
where the wave vector $\kk(t)$ and the phase $\varphi(t)$ change randomly
from one time step to the next.
This forcing function is therefore white noise in time and consists of plane
waves with average wavenumber $\kf$ such that $|\kk|$ lies in an interval
$\kf-\delta k/2\leq|\kk|<\kf+\delta k/2$ of width $\delta k$.
Here, ${\cal N}=f_0/\delta t^{1/2}$ is a normalization factor, where
$\delta t$ is the time step and $f_0$ is varied to achieve a certain
magnetic field strength after a certain time, and
$\tilde{\ff}({\kk})=(\kk\times\ee)/[\kk^2-(\kk\cdot\ee)^2]^{1/2}$
is a nonhelical forcing function.
Here, $\ee$ is an arbitrary unit vector that is not aligned with $\kk$.
Note that $|\ff|^2=1$.
Following earlier work, the forcing is only enabled during the
time interval $1\leq t\leq 2$. 
The kinetic and magnetic energy densities are defined as
$\EEK(t) = \braket{\rho\mathbf{u}^2}/2$ and
$\EEM(t) = \braket{\mathbf{B}^2}/2$, respectively. 

\begin{figure*}[t!]\begin{center}
\includegraphics[width=.99\columnwidth]{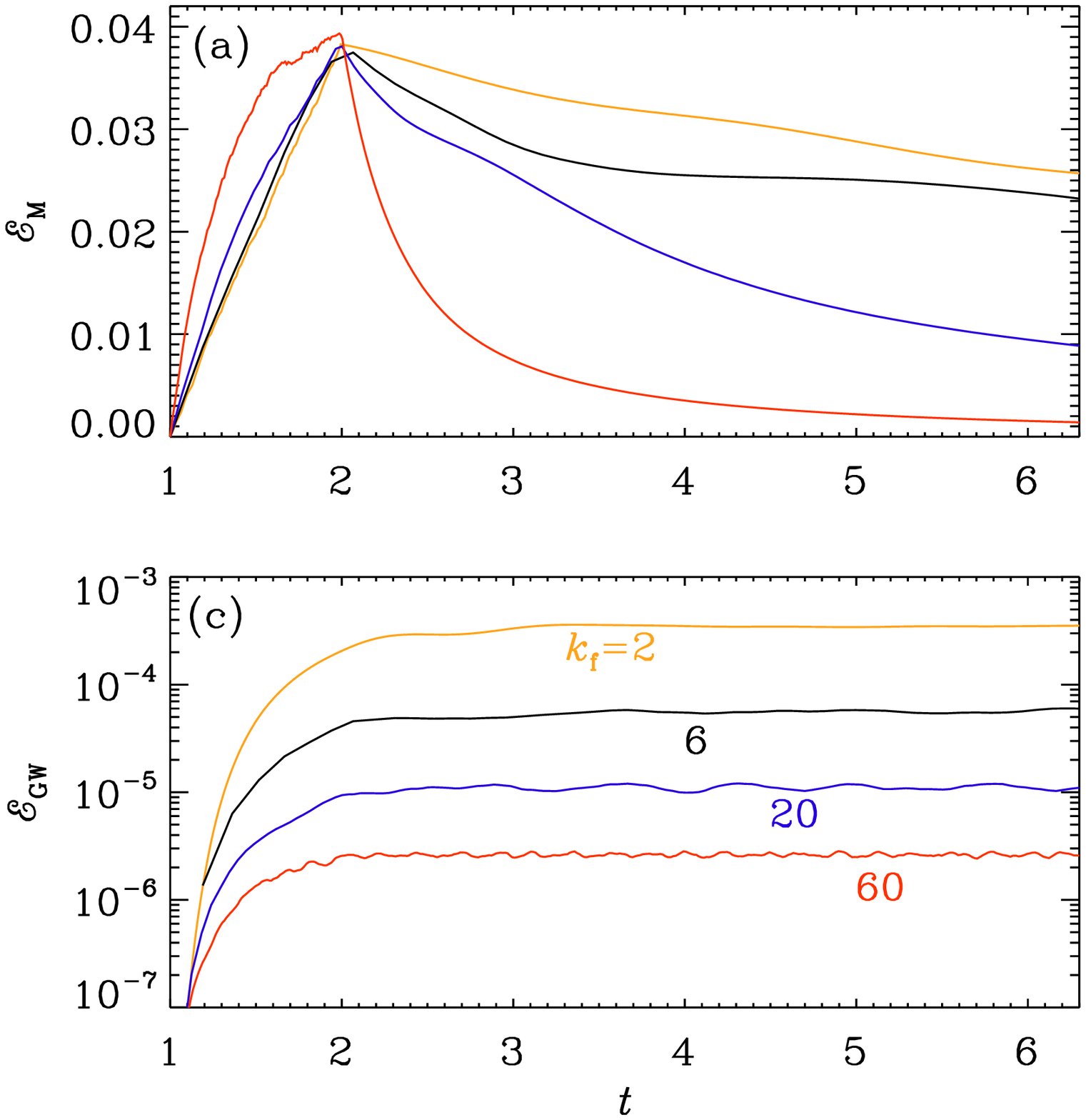}
\includegraphics[width=.99\columnwidth]{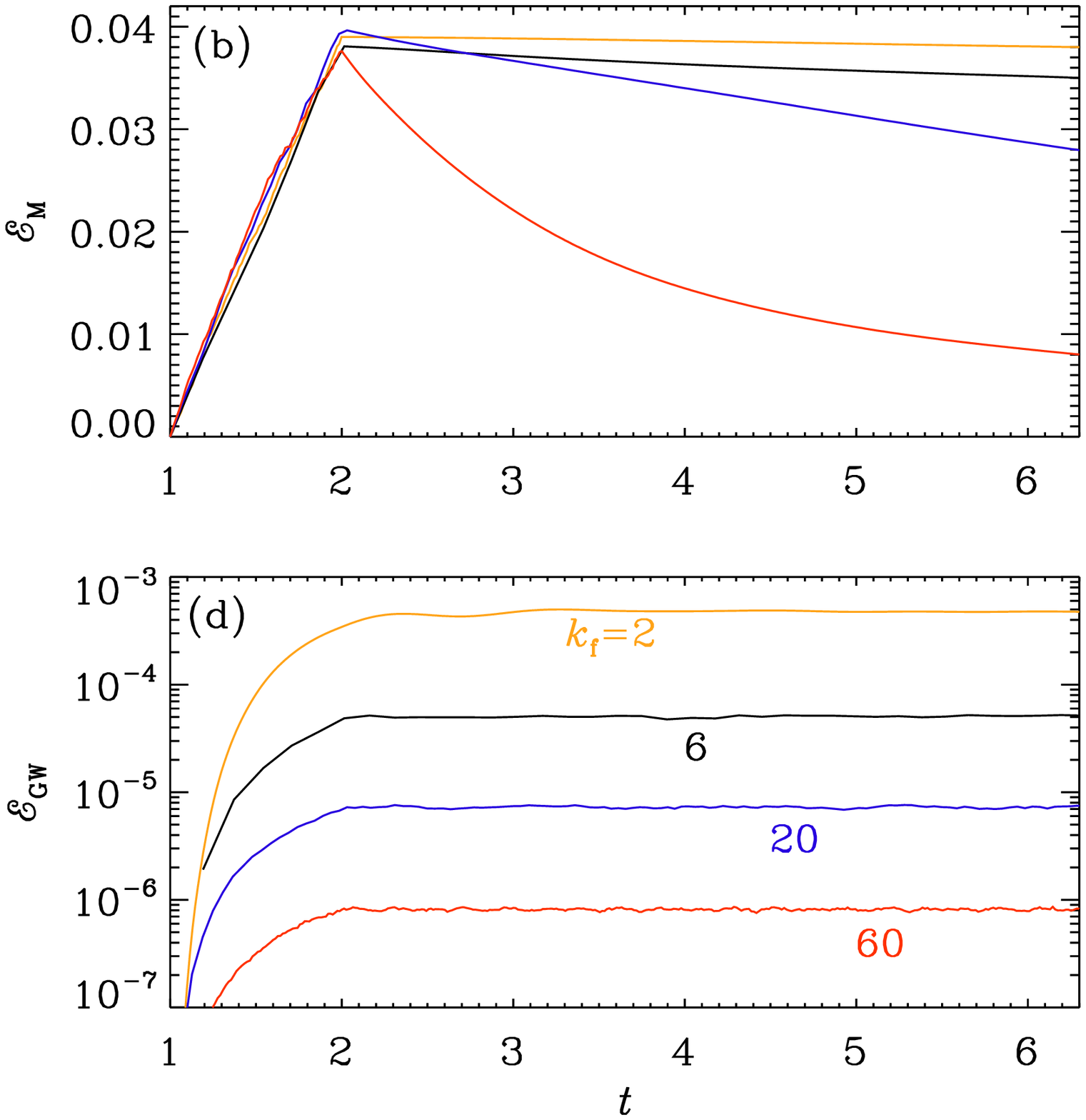}
\end{center}\caption[]{ 
Evolution of (a,b) $\EEM(t)$ and (c,d) $\EEGW(t)$ for
(a,c) nonhelical and (b,d) helical cases.
Orange, black, blue, and red are for $\kf=2$, 6, 20, and 60, respectively.
}\label{pcomp512}\end{figure*}

The vigor of turbulence is characterized by the Reynolds number,
$\Rey=\urms/\nu\kf$, where $\urms$ is the maximum rms velocity.
It can only be determined a posteriori from the velocity resulting
from the magnetic field through the Lorentz force.
For all our runs, we use $\eta=\nu$.

\subsection{Turbulent Decay Laws}

Turbulence is known to decay in power-law fashion \cite{BP56,Saf67} such that
the magnetic energy $\EEM(t)$ decays with time $t$ like $\Delta t^{-p}$
and the correlation length $\xiM(t)$ increases like $\Delta t^{q}$,
where $\Delta t=t-t_{\rm off}$ is the time interval after the forcing
has been turned off.
The exponents $p$ and $q$ are positive and depend on the physical
circumstances (magnetically or kinetically dominated turbulence),
and whether or not there is magnetic helicity.
In helical turbulence, for example, one finds $p=q=2/3$, while for
non-helical magnetically dominated turbulence one finds $p=1$ and
$q=1/2$, although other variants are sometimes possible
\cite{Brandenburg:2017rnt,Hosking21}.

In this paper, we are specifically interested in the dependence of
the decay behavior on the forcing wave number $\kf$ of the turbulence
while it was still being driven.
The parameter $\kf$ enters through the prefactor in the decay law.

Furthermore, $\Delta t^{-p}$ would become infinite for $p>0$
and $\Delta t=0$ (when $t=t_{\rm off}$).
The singularity of $\Delta t^{-p} $ at $t=0$ is a consequence of a
simplified description at the initial time moment. 
For this reason, it is convenient to express the decay laws as
\begin{equation}
\EEM(t)=\EEM^{\max}\,(1+\Delta t/\tau)^{-p},
\label{EEMt}
\end{equation}
where $\tau$ is the turnover time, which we will treat as an empirical
parameter that we expect to be of the order of $(\vA\kf)^{-1}$, where
$\vA=(3\EEM^{\max}/2)^{1/2}$ is the Alfv\'en speed, evaluated
at the time when $\EEM$ reaches its maximum value $\EEM^{\max}$.
In some simulations of purely hydrodynamic turbulence, we replace
$\EEM$ by $\EEK$ in \Eq{EEMt} and use $\tau=(\urms\kf)^{-1}$ with
$\urms=(2\EEK)^{1/2}$ as the nominal turnover time. 
We recall here that we are using nondimensional variables where the
radiation energy density is unity.

For all our simulations, we choose $t_{\rm off}=2$, i.e., turbulence
is being driven for one Hubble time during $1\leq t\leq2$.
In the following, we vary $\kf$ between 2 and 60.
For $\kf=60$, we find that $\tau$ is shorter than a Hubble time,
but in all other cases, it exceeds it by up to factors between
ten (in the nonhelical cases) and a hundred (in helical cases).

We arrange the strength of the forcing $f_0$ such that $\EEM$ is similar
for different values of $\kf$.
This allows us then to determine the resulting GW energy solely
as a function of $\kf$.
For small values of $\kf$, the turbulence may not be able to reach
a statistically steady state by the time $t_{\rm off}$, when the
driving is turned off.

It is therefore necessary to adjust $f_0$ for each value of $\kf$
separately.
Once we have two values of $\EEM^{\max}$ that are close enough
to the target strength, we determine the desired forcing
strength through linear interpolation.
We also consider the case of different values of $f_0$ for a fixed value
of $\kf$ (Runs~noh5,6 and Runs~hel5,6).

\section{Numerical simulations}
\label{NumericalSimulations}

We solve the governing equations using the {\sc Pencil Code} \cite{PC},
where the GW solver has already been implemented \cite{Pol:2018pao}. 
We consider a cubic domain of side length $2\pi/k_1$, where $k_1$ is
the smallest wave number in the domain.
We choose $k_1=\kf/6$, so that the scale separation between the initial
spectral peak and the lowest wave number in the domain is six. 
In the following, we discuss the results for different values of $\kf$.
The temporal growth of $\EEM(t)$ is similar for small values of $\kf$;
see \Fig{pcomp512}(a) and (b), where we compare the evolution of $\EEM$ and
$\EEGW$ for the nonhelical and helical cases.
The parameters of those runs are listed in \Tabs{Tsummary512}{Tsummary512hel}
(for nonhelical and helical runs).
The numerical resolution is $512^3$ mesh points, except for run noh1,
where we use $1024^3$ mesh points.
Unless specified otherwise, we use $\nu=\eta=5\times10^{-5}$.

\begin{table*}[t!]\caption{
Summary of runs with nonhelical turbulence.
}\vspace{12pt}\centerline{\begin{tabular}{ccccccccc|ccc}
Run & $\kf$ & $k_1$ & $f_0$ & $p$ & $\tau$ & $\EEM^{\max}$ & $\EEGW^{\rm sat}$ & $\hrms^{\rm sat}$ &
$B$ [$\mu$G] & $h_0^2\OmGW(f)$ & $h_c$ \\
\hline
noh1&$   2$&$ 0.3$&$1.9\times10^{-1}$&$ 1.0$&$  16$&$3.83\times10^{-2}$&$3.53\times10^{-4}$&$4.83\times10^{-2}$&$0.78$&$1.09\times10^{-8}$&$4.83\times10^{-14}$\\
noh2&$   6$&$   1$&$6.0\times10^{-2}$&$ 1.0$&$ 4.5$&$3.75\times10^{-2}$&$5.61\times10^{-5}$&$7.06\times10^{-3}$&$0.78$&$1.73\times10^{-9}$&$7.07\times10^{-15}$\\
noh3&$  20$&$   3$&$2.3\times10^{-2}$&$ 1.3$&$ 2.0$&$3.81\times10^{-2}$&$1.11\times10^{-5}$&$1.15\times10^{-3}$&$0.78$&$3.44\times10^{-10}$&$1.15\times10^{-15}$\\
noh4&$  60$&$  10$&$1.0\times10^{-2}$&$ 1.4$&$0.43$&$3.93\times10^{-2}$&$2.62\times10^{-6}$&$1.65\times10^{-4}$&$0.79$&$8.10\times10^{-11}$&$1.65\times10^{-16}$\\
\hline
noh5&$   2$&$ 0.3$&$1.0\times10^{-1}$&  --- &  --- &$1.06\times10^{-2}$&$2.70\times10^{-5}$&$1.40\times10^{-2}$&$0.41$&$8.37\times10^{-10}$&$1.40\times10^{-14}$\\
noh6&$   2$&$ 0.3$&$3.0\times10^{-1}$&  --- &  --- &$9.48\times10^{-2}$&$2.08\times10^{-3}$&$1.02\times10^{-1}$&$ 1.2$&$6.42\times10^{-8}$&$1.02\times10^{-13}$\\
noh7&$   6$&$   1$&$2.0\times10^{-2}$&  --- &  --- &$4.63\times10^{-3}$&$6.56\times10^{-7}$&$8.10\times10^{-4}$&$0.27$&$2.03\times10^{-11}$&$8.11\times10^{-16}$\\
noh8&$   6$&$   1$&$1.0\times10^{-1}$&  --- &  --- &$8.90\times10^{-2}$&$3.89\times10^{-4}$&$1.67\times10^{-2}$&$ 1.2$&$1.20\times10^{-8}$&$1.67\times10^{-14}$\\
\label{Tsummary512}\end{tabular}}\end{table*}

\begin{table*}[t!]\caption{
Similar to \Tab{Tsummary512}, but for helical turbulence.
}\vspace{12pt}\centerline{\begin{tabular}{ccccccccc|ccc}
Run & $\kf$ & $k_1$ & $f_0$ & $p$ & $\tau$ & $\EEM^{\max}$ & $\EEGW^{\rm sat}$ & $\hrms^{\rm sat}$ &
$B$ [$\mu$G] & $h_0^2\OmGW(f)$ & $h_c$ \\
\hline
hel1&$   2$&$ 0.3$&$1.9\times10^{-1}$&$0.67$&$ 100$&$3.90\times10^{-2}$&$4.85\times10^{-4}$&$4.33\times10^{-2}$&$0.79$&$1.50\times10^{-8}$&$4.33\times10^{-14}$\\
hel2&$   6$&$   1$&$5.6\times10^{-2}$&$0.67$&$  20$&$3.81\times10^{-2}$&$5.05\times10^{-5}$&$4.69\times10^{-3}$&$0.78$&$1.56\times10^{-9}$&$4.69\times10^{-15}$\\
hel3&$  20$&$   3$&$2.0\times10^{-2}$&$0.67$&$ 4.0$&$3.96\times10^{-2}$&$7.26\times10^{-6}$&$6.66\times10^{-4}$&$0.80$&$2.24\times10^{-10}$&$6.66\times10^{-16}$\\
hel4&$  60$&$  10$&$6.5\times10^{-3}$&$0.67$&$0.50$&$3.76\times10^{-2}$&$8.15\times10^{-7}$&$7.18\times10^{-5}$&$0.78$&$2.52\times10^{-11}$&$7.18\times10^{-17}$\\
\hline
hel5&$   2$&$ 0.3$&$1.0\times10^{-1}$&  --- &  --- &$1.06\times10^{-2}$&$3.61\times10^{-5}$&$1.08\times10^{-2}$&$0.41$&$1.12\times10^{-9}$&$1.08\times10^{-14}$\\
hel6&$   2$&$ 0.3$&$3.0\times10^{-1}$&  --- &  --- &$9.85\times10^{-2}$&$3.07\times10^{-3}$&$1.12\times10^{-1}$&$ 1.3$&$9.49\times10^{-8}$&$1.12\times10^{-13}$\\
hel7&$   6$&$   1$&$2.0\times10^{-2}$&  --- &  --- &$4.93\times10^{-3}$&$8.33\times10^{-7}$&$6.26\times10^{-4}$&$0.28$&$2.58\times10^{-11}$&$6.26\times10^{-16}$\\
hel8&$   6$&$   1$&$1.0\times10^{-1}$&  --- &  --- &$1.20\times10^{-1}$&$5.09\times10^{-4}$&$1.59\times10^{-2}$&$ 1.4$&$1.57\times10^{-8}$&$1.59\times10^{-14}$\\
\label{Tsummary512hel}\end{tabular}}\end{table*}

In \Tab{Tsummary512}, we have quoted the values of $\EEGW^{\rm sat}$ and
$\hrms^{\rm sat}$ obtained at the end of the simulation.
To compute the relic observable $h_0^2\OmGW$ at the present time,
we have to multiply $\EEGW^{\rm sat}$ by a factor
$(H_\ast/H_0)^2 (a_*/a_0)^4$;
see Refs.~\cite{Pol:2019yex,Pol:2018pao} for details. 
Using $g_*=15$ and $T_*=150\MeV$, we find $H_*=1.8\times10^4\s^{-1}$, 
and thus this factor is $\approx3\times10^{-5}$.
The largest value of $\EEGW^{\rm sat}$ quoted in
\Tab{Tsummary512} is $3.5\times10^{-4}$ and corresponds 
therefore to $h_0^2\OmGW\approx10^{-8}$.
Likewise, the values of $\hrms^{\rm sat}$ in \Tab{Tsummary512} 
have to be multiplied by $a_0^{-1}\approx10^{-12}$ to
obtain the observable $h_c$ at the present time; see \Eq{Dfactor}.
Again, the largest value of $\hrms^{\rm sat}=5\times10^{-2}$ 
corresponds therefore to the observable $h_c=5\times10^{-14}$.

\begin{figure*}[t!]\begin{center}
\includegraphics[width=.99\columnwidth]{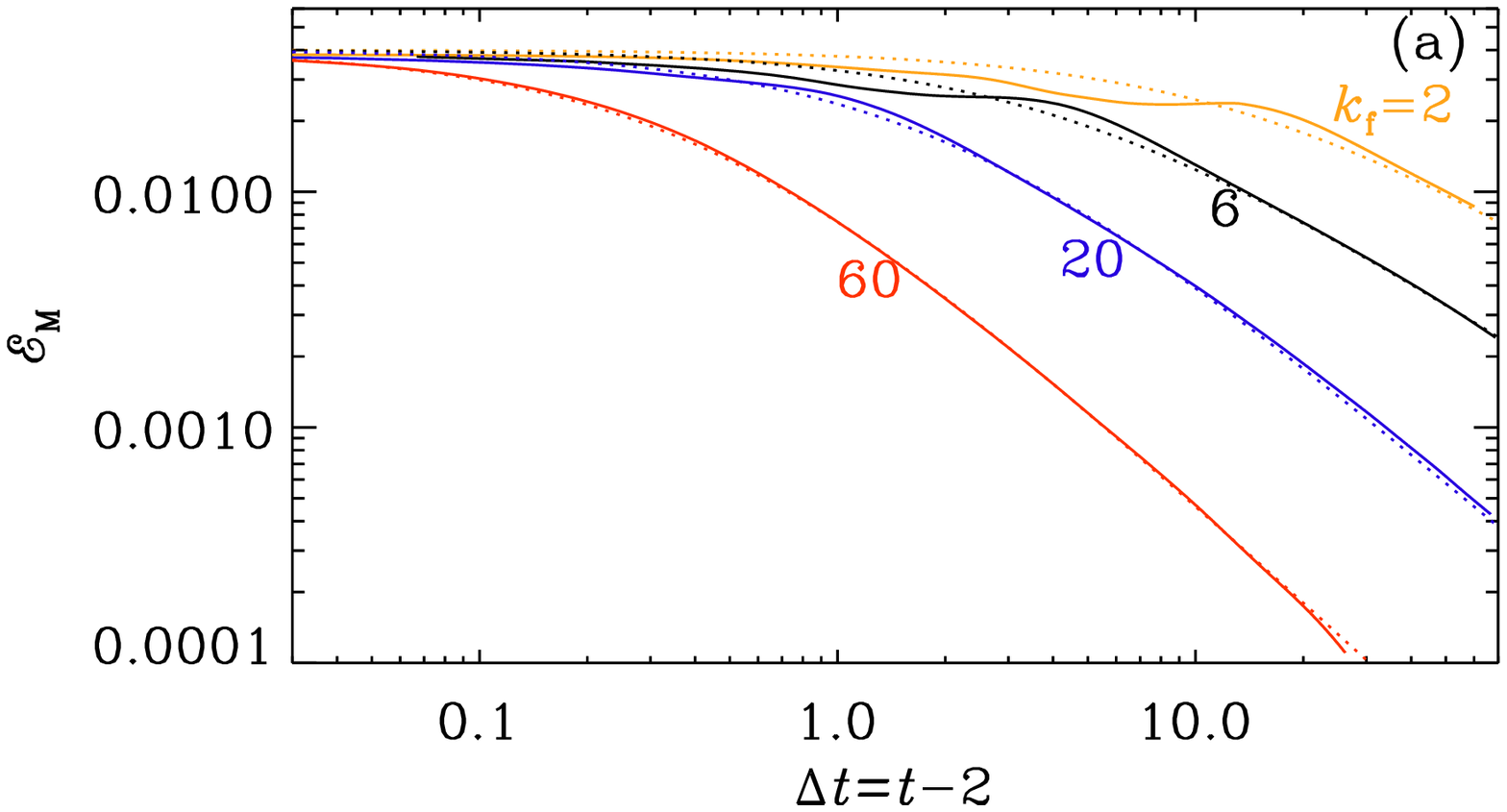}
\includegraphics[width=.99\columnwidth]{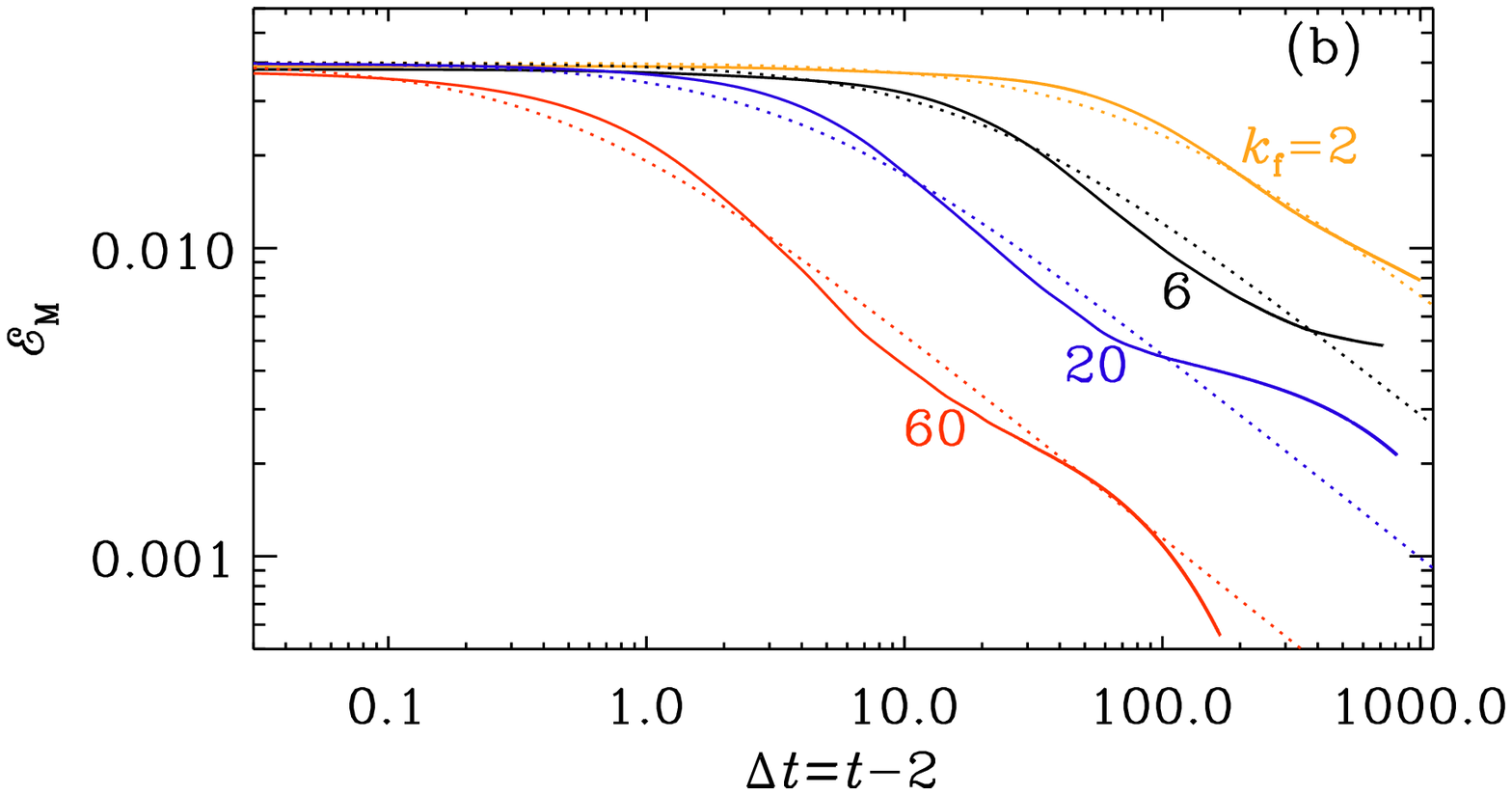}
\end{center}\caption[]{
Similar to \Fig{pcomp512}(a) and (b), but in a double-logarithmic representation 
for (a) nonhelical and (b) helical cases, 
where $\EEM$ is now plotted versus $\Delta t\equiv t-2$, the time after
which the electromagnetic source is turned off.
}\label{pcomp512_oo}\end{figure*}

\begin{table*}[t!]\caption{
Comparison of nonhelical magnetic turbulence (mag) with irrotational (irro)
and vortical (vort) turbulence.
}\vspace{12pt}\centerline{\begin{tabular}{lccccc|ccc}
Type & $f_0$ & $\nu$ & $\EEM^{\max}$ & $\EEGW^{\rm sat}$ & $\hrms^{\rm sat}$ &
$B$ [$\mu$G] & $h_0^2\OmGW(f)$ & $h_c$ \\
\hline
magnetic&$1.9\times10^{-1}$&$5.0\times10^{-5}$&$3.83\times10^{-2}$&$3.53\times10^{-4}$&$4.83\times10^{-2}$&$0.78$&$1.09\times10^{-8}$&$4.83\times10^{-14}$\\
vortical&$3.8\times10^{-1}$&$1.0\times10^{-2}$&$4.21\times10^{-2}$&$8.81\times10^{-4}$&$8.26\times10^{-2}$&$0.82$&$2.73\times10^{-8}$&$8.27\times10^{-14}$\\
irrotational&$7.0\times10^{-1}$&$2.0\times10^{-2}$&$4.26\times10^{-2}$&$8.30\times10^{-4}$&$7.95\times10^{-2}$&$0.83$&$2.57\times10^{-8}$&$7.96\times10^{-14}$\\
\label{Tsummary_hydro}\end{tabular}}\end{table*}

To simplify the comparisons, we have arranged 
the forcing amplitude $f_0$ such that
$\EEM^{\max}$ is similar in certain cases.
The values of $\EEM^{\max}$ listed in the upper block of
\Tabs{Tsummary512}{Tsummary512hel} (for nonhelical and helical
hydromagnetic turbulence, respectively) are around 0.038 and
correspond to $0.8\uG$.
The growth phase of $\EEM(t)$ is similar, but the decay is significantly
slower when $\kf$ is smaller.
The GW energy saturates at a value $\EEGW^{\rm sat}$ some time after
$\EEM(t)$ has reached its maximum, and is smaller for larger values
of $\kf$.

It is important to realize that in all four cases, the decay of the 
magnetic energy follows an approximate power law decay, as given
by \Eq{EEMt}.
To see this, we the plot in \Fig{pcomp512_oo} the evolution of $\EEM$
versus $t-2$ in a double-logarithmic representation.
The parameters $p$ and $\tau$ describe the
decay and are also listed in \Tab{Tsummary512}.

\begin{figure*}[t!]\begin{center}
\includegraphics[width=\columnwidth]{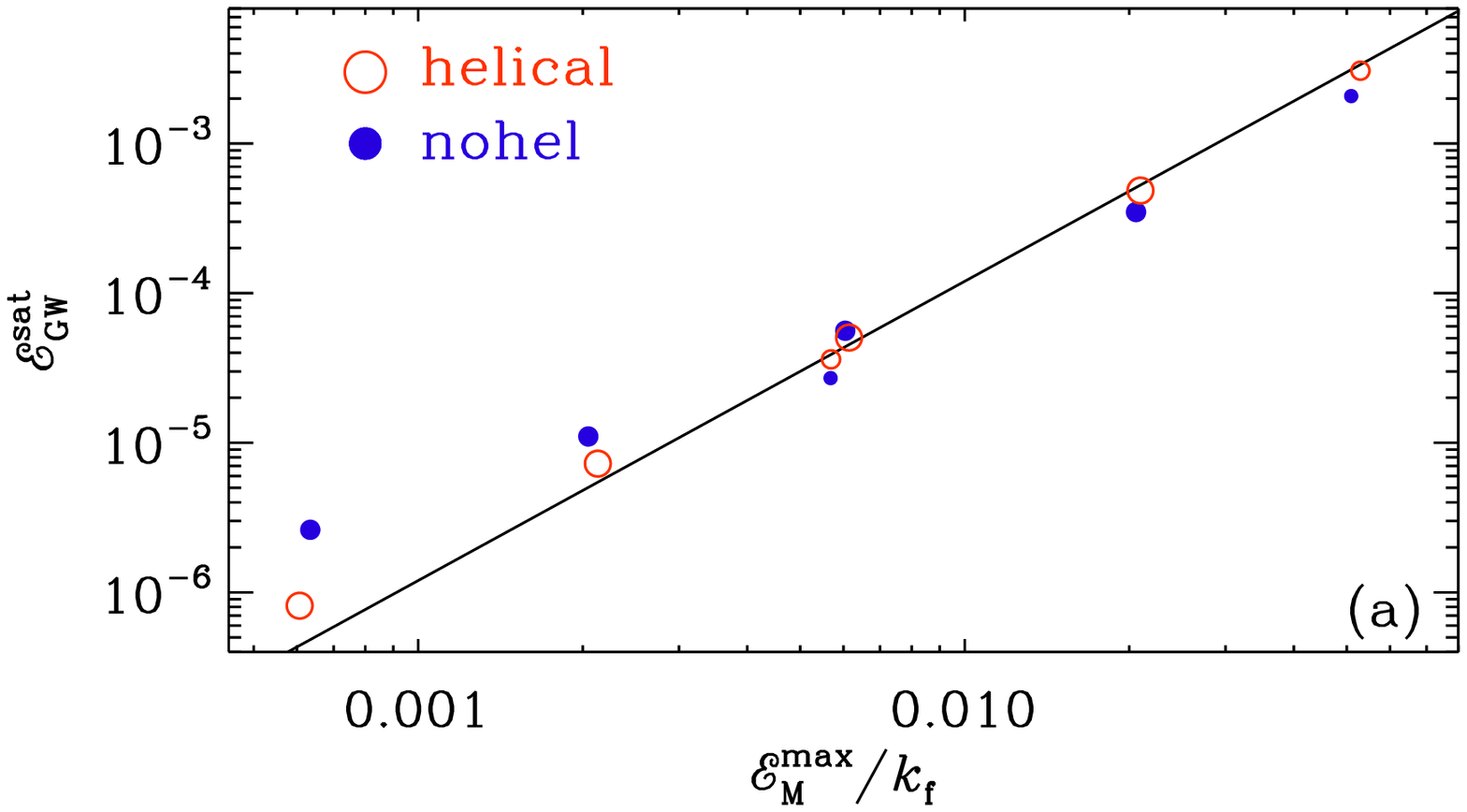}
\includegraphics[width=\columnwidth]{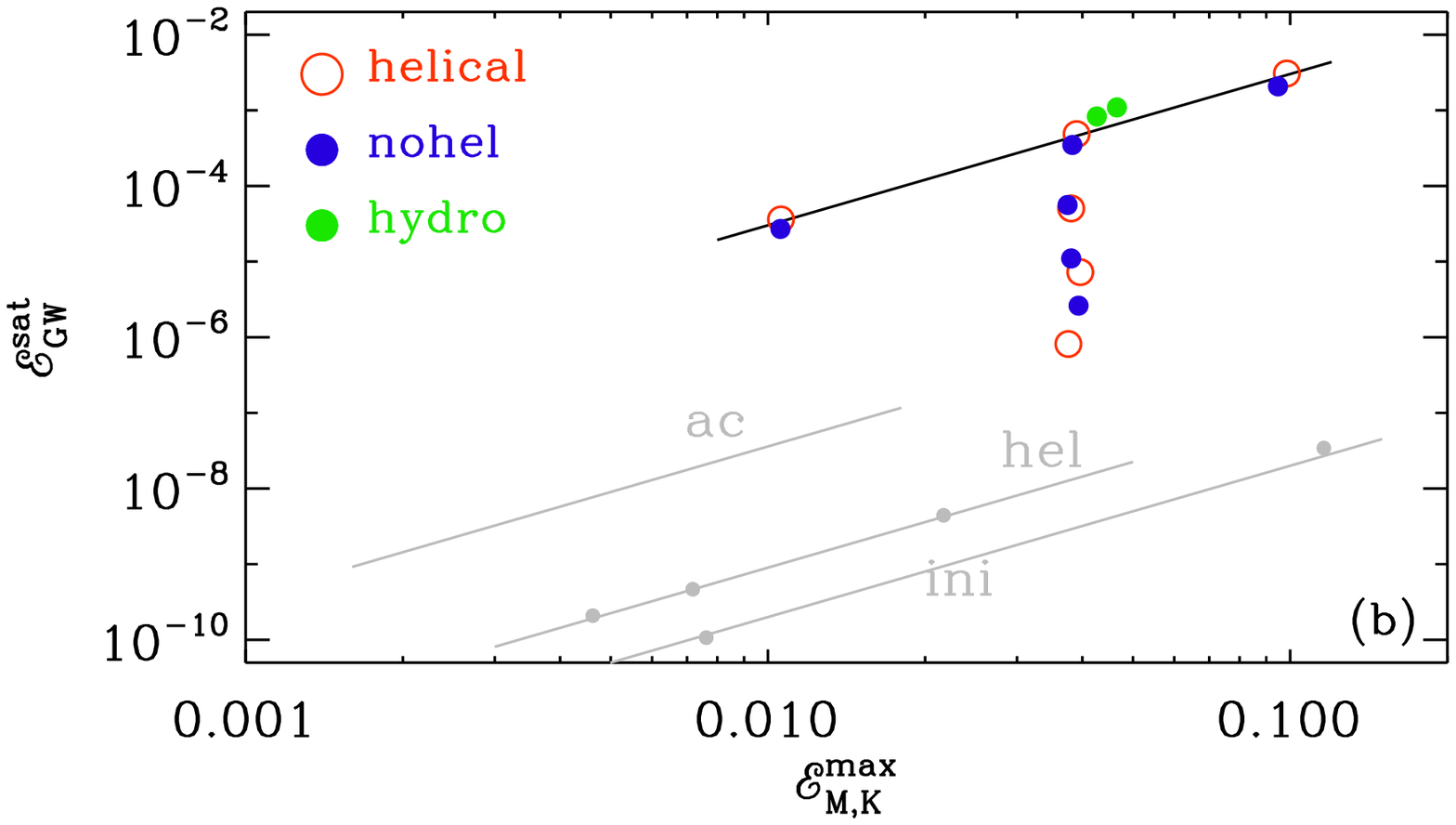}
\end{center}\caption[]{
(a) $\EEGW$ versus $\EEM/\kf$; the straight line shows
$\EEGW=5.2\times10^{-4}\,(\EEM/\kf)^{1/2}$.
(b) Positions of our runs in a diagram showing
$\EEGW^{\rm sat}$ versus $\EEM^{\max}$.
For orientation the old data points of the Ref.~\cite{Pol:2019yex}
are shown as gray symbols.
The open red (filled blue) symbols are for the helical (nonhelical) runs.
The green symbols refer to the two hydromagnetic runs of 
\Tab{Tsummary_hydro}.
}\label{pres}\end{figure*}

\begin{figure*}[t!]\begin{center}
\includegraphics[width=.99\columnwidth]{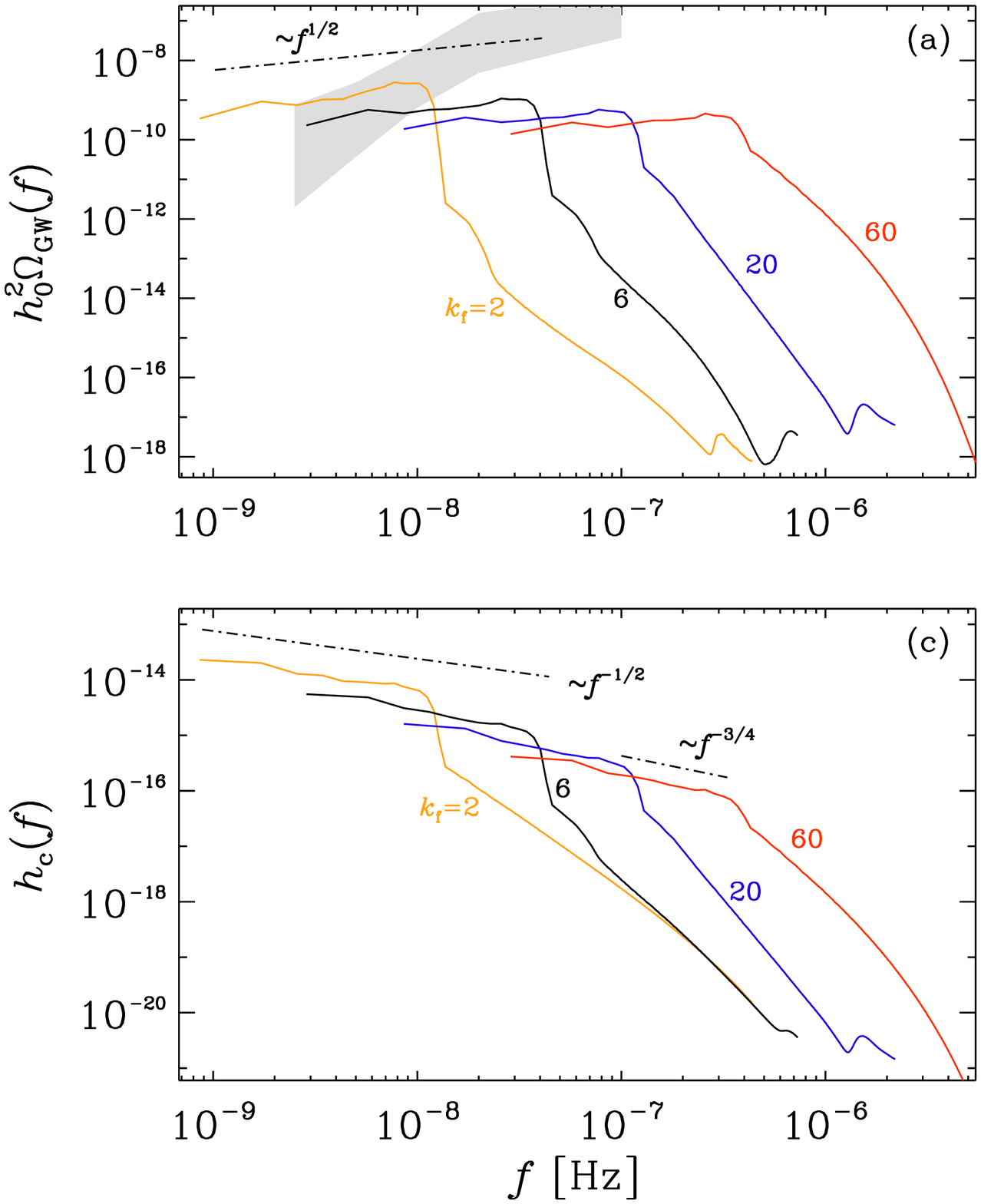}
\includegraphics[width=.99\columnwidth]{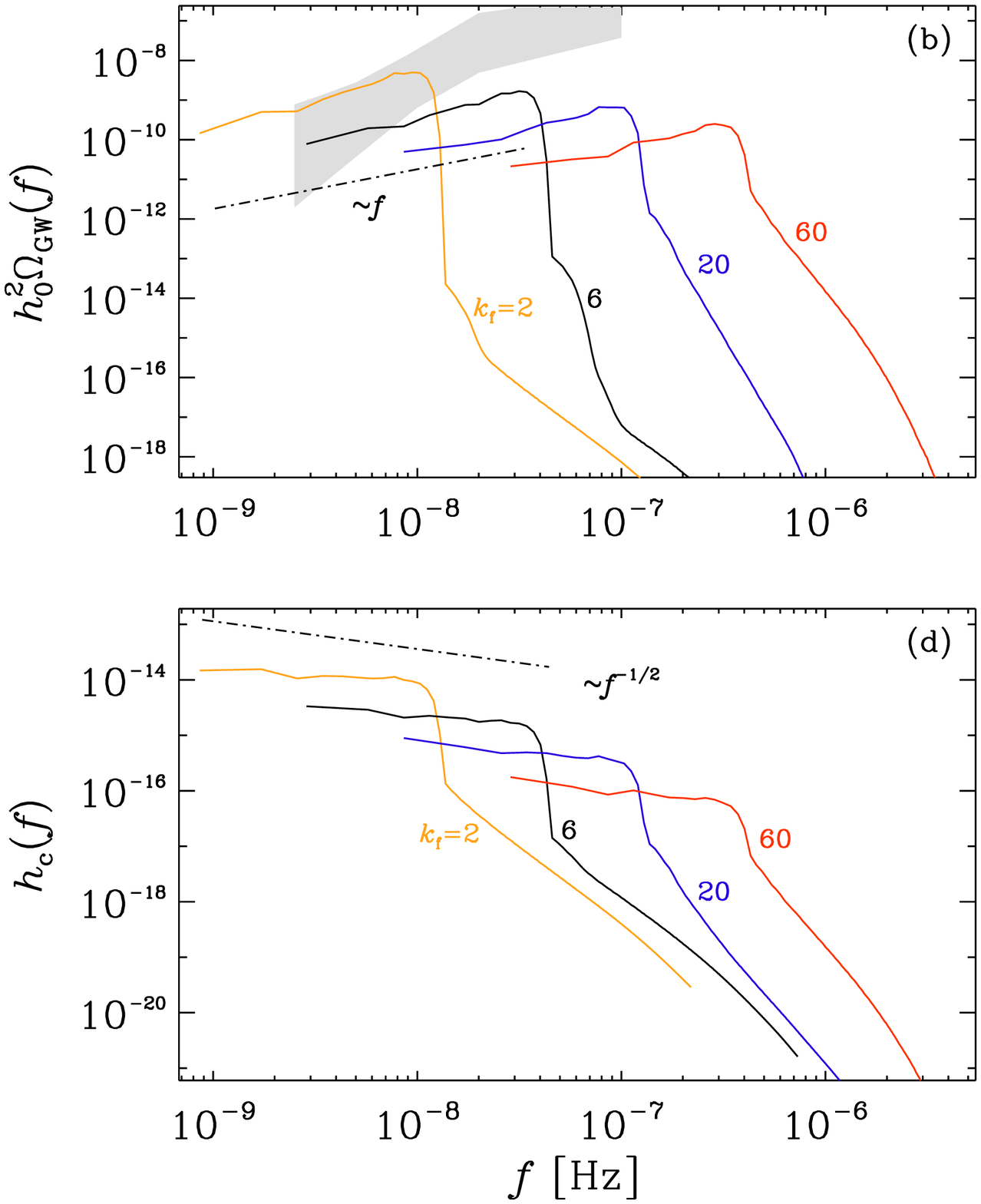}
\end{center}\caption[]{
(a,b) $h_0^2\OmGW(f)$ and (c,d) $h_c(f)$ at the present time for all four runs 
presented in \Tab{Tsummary512}, for the
(a,c) nonhelical and (b,d) helical runs. 
The $2\sigma$ confidence contour for the 30-frequency power law of
the NANOGrav 12.5-year data set is shown in gray.
}\label{pspecm}\end{figure*}

\begin{figure*}[t!]\begin{center}
\includegraphics[width=\textwidth]{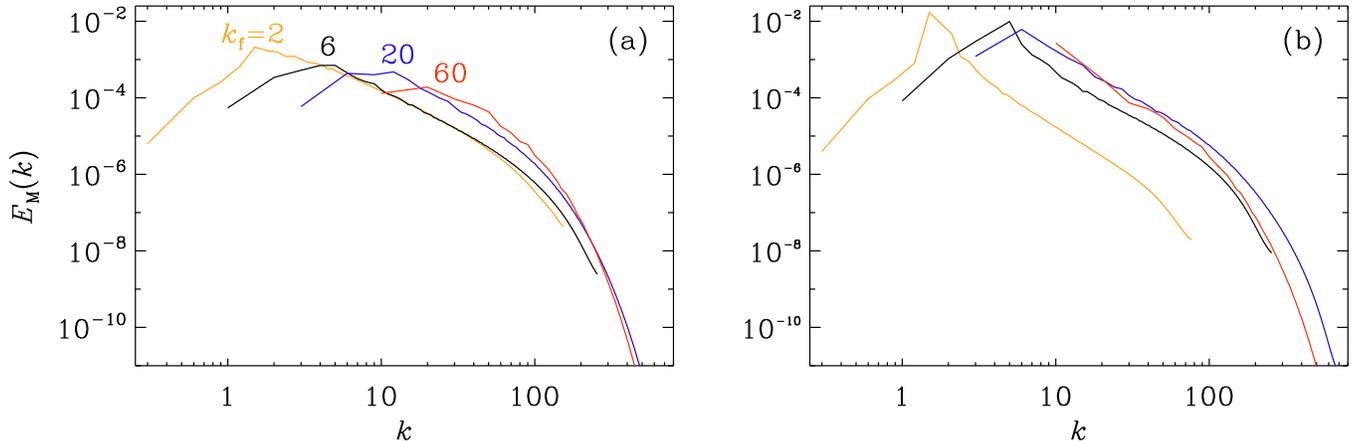}
\end{center}\caption[]{
Magnetic energy spectra for the (a) nonhelical and (b) helical cases.
}\label{pbspecm}\end{figure*}

\begin{figure}[t!]\begin{center}
\includegraphics[width=\columnwidth]{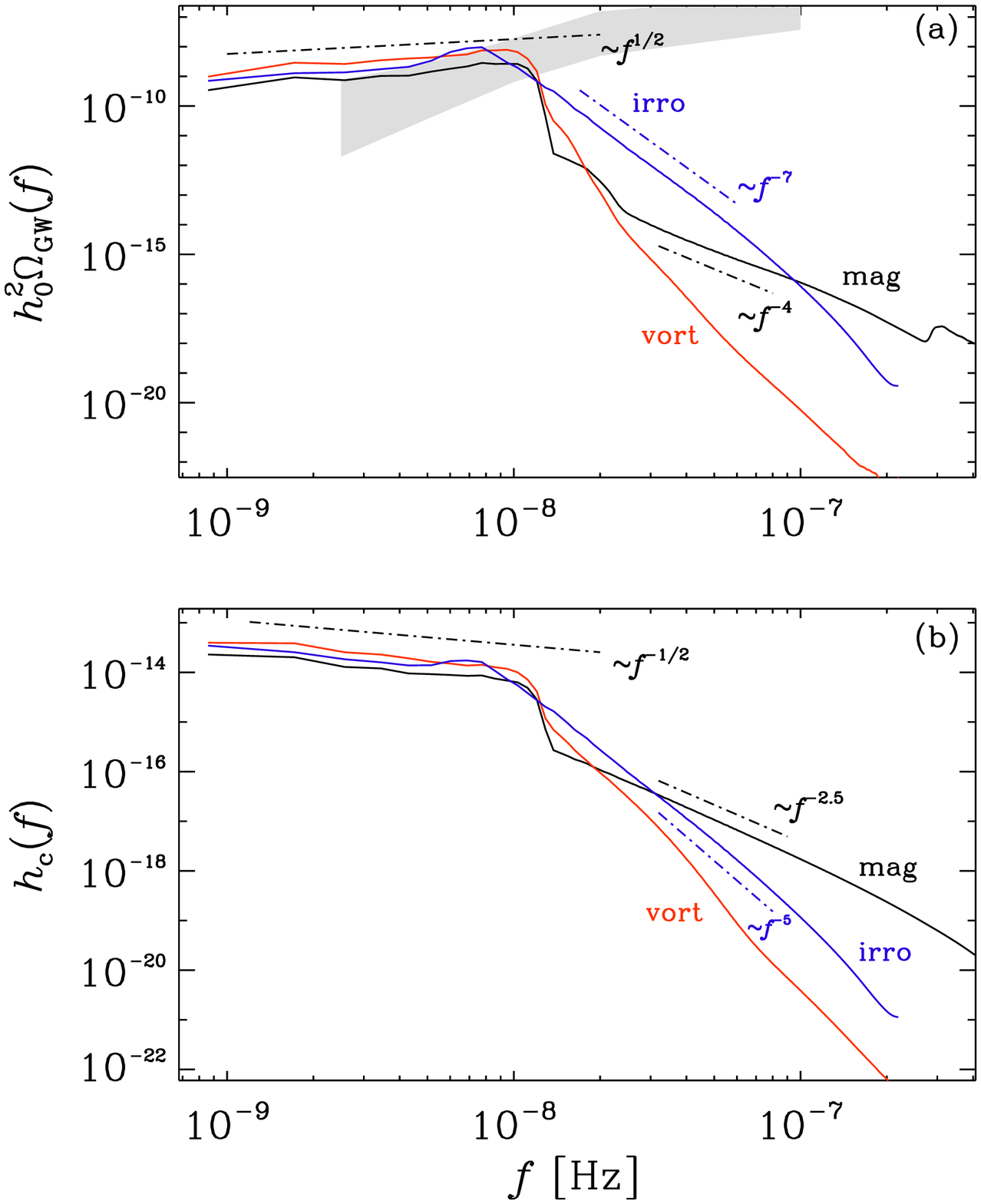}
\end{center}\caption[]{
Similar to \Fig{pspecm}, showing (a) $h_0^2\OmGW(f)$ and (b) $h_c(f)$,
but comparing vortical (red) and 
irrotational turbulence (blue) with MHD turbulence (black).
}\label{pspecm_other}\end{figure}

Our results confirm that the turbulence decays more slowly for
large values of $\tau$, or small values of $\kf$. 
As already found from earlier simulations \cite{Pol:2019yex},
the GW energy generally decreases with increasing $\kf$.
This is seen more clearly in a diagram of $\EEGW$ versus $\EEM/\kf$;
see \Fig{pres}(a).

For $\kf=2$, we have performed additional simulations with smaller
and larger values $f_0$, both with and without helicity.
The resulting values of $\EEGW^{\rm sat}$ obey quadratic scaling
of the form
\begin{equation}
\EEGW^{\rm sat}=\left(q\EEM^{\max}/\kf\right)^2
\label{qdef}
\end{equation}
with a coefficient $q=1.1$; see the straight line in \Fig{pres}(a).
Only the data point for $\kf=60$ is slightly above the line
represented by \Eq{qdef}.
This could be an artefact of our Reynolds numbers still not being large
enough in our simulations, especially for large value of $\kf$.

To compare with earlier work, we show in \Fig{pres}(b) the positions
of our runs in a $\EEGW^{\rm sat}$ versus $\EEM^{\max}$ diagram.
For orientation, we also show the data points from Ref.~\cite{Pol:2019yex}.
We see that the new data points are well above the older ones of
Ref.~\cite{Pol:2019yex}. 
This is mainly a consequence of using here smaller values of $\kf$
(2--60, compared to 600 in Ref.~\cite{Pol:2019yex}).
For $\kf=2$ and $k_1=0.3$, we show here the results for 
hydrodynamic runs using irrotational and vortical forcings;
see the green symbols in \Fig{pres}(b).
Those runs are listed in \Tab{Tsummary_hydro} and compared with 
the nonhelical magnetic turbulence run `noh1'.

In \Fig{pspecm}, we plot the resulting present-day GW energy and strain
spectra for our four runs with $\kf=2$, 6, 20, and 60, both without and
with helicity in the driving function $\FFF$.
The first two cases with $\kf=2$ and 6 lie well within the frequency
and amplitude range accessible to NANOGrav.
In all cases, the spectra show a sharp drop slightly above the peak
frequency.
This is a consequence of the rapid temporal growth of the spectra,
which leads to a correspondingly large growth at the peak frequency,
while at higher frequencies, the spectrum settled at values that were
determined by somewhat earlier times when the energy was still weaker.

At frequencies below the peak, we now find a spectrum that is even
shallower than the $h_0^2\Omega_{\rm GW}(f)\propto f$ spectrum found
already earlier \cite{Pol:2019yex}.
A spectrum shallower than proportional to $f$, such as the present
$f^{1/2}$ spectrum, could perhaps be explained by the finite size of the
computational domain; see Ref.~\cite{Brandenburg:2021aln}, who found even
a $f^{-1/2}$ spectrum for $\Omega_{\rm GW}(f)$. 
Alternatively, the shallower spectrum might well be physical,
or at least significantly extended over a substantial frequency
interval below the peak frequency, for example due to inverse 
cascading in helical \cite{FPLM75} and nonhelical \cite{BK17} cases.

In the absence of sources, a $\Omega_{\rm GW}(f)\propto f^\alpha$ spectrum
implies $h_c(f)\propto f^{\alpha/2-1}$ for arbitrary spectral indices $\alpha$.
For $\alpha=1/2$, we would thus expect $h_c(f)\propto f^{-3/4}$.
However, the observed strain spectrum, $h_c(f)\propto f^{-1/2}$, seems to agree
with that found previously from numerical simulations \cite{Pol:2019yex}. 
However, looking more carefully at the strain spectrum for $\kf=60$, we see
a $h_c(f)\propto f^{-3/4}$ spectrum is actually compatible with the simulation;
see the corresponding dashed-dotted line in \Fig{pspecm}(c).
This agreement is probably related to the fact that the turnover time is
shorter for the run with $\kf=60$, compared with those at smaller values 
(i.e., longer turbulence driving time will allow for
more efficient inverse cascading).

In the runs with helicity, we do find $h_0^2\Omega_{\rm GW}(f)\propto f$,
together with a slight enhancement just before reaching a maximum.
The subsequent decay for larger values of $f$ is much steeper in the
case with helicity than without.
Furthermore, in $h_c(f)$ we see a sharper drop to the right of the 
maximum than in simulations without helicity.
These differences in the spectra for helical and nonhelical cases
are surprisingly strong and might allow us to infer the presence
of magnetic helicity once such a spectrum is detected. 

It is important to note that the $h_0^2\Omega_{\rm GW}\propto f$ spectra
in Figs.~\ref{pres}(a) and (c) show an increase towards smaller $\kf$.
This is to be expected from \Eq{qdef}, but it was not included in the
sketch of Ref.~\cite{Neronov:2020qrl}; see their Fig.~(1).
By contrast, in their Eq.~(4), an effectively cubic dependence on
the magnetic energy was motivated.

The underlying magnetic energy spectrum is shown in \Fig{pbspecm}(a) for
nonhelical and in \Fig{pbspecm}(b) helical cases where $\kf$ is ranging from 2 to 60.
Those are averaged spectra obtained by averaging over the time interval
$15\leq t\leq20$.
In the nonhelical case, the amplitude of the spectrum is smaller for
larger values of $\kf$, because here the energy has decayed more rapidly.
In the helical case, the spectra have approximately the same height for
all values of $\kf$.
This is because the height of the spectrum
is related to the helicity, which is conserved.
For small values of $\kf$, the spectrum has a more extended subinertial
range.
This is because the turnover time is larger and there was not enough
time for the inverse cascade to produce energy and small values of $k$.

\begin{figure*}[t!]
\begin{center}
\includegraphics[scale=1]{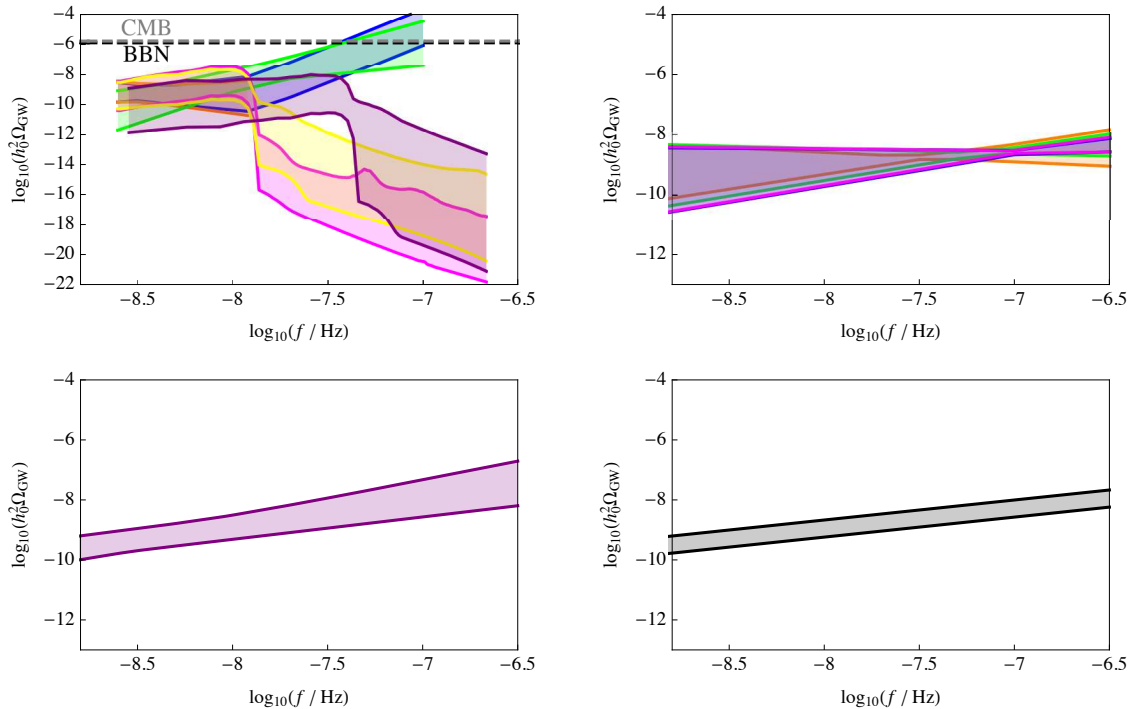}
\end{center}
\caption[]
{
{Upper left: Runs noh5,6 (yellow) and hel5,6 (magenta),
which correspond to $\kf = 2$, and noh7,8 (purple), corresponding to $\kf = 6$,
compared with the NANOGrav 12.5-year 2$\sigma$ contours (as shown in Fig.~\ref{fig:NGcontours}). 
Additionally, the gray and black dashed horizontal lines show the CMB and BBN {\it integrated} bounds on $h_0^2\OmGW(f)$ (see \cite{MaggioreText2} for details).} 
{Upper right: Contours representing four different cosmic string average power spectrum
models with different tensions, as described in \cite{Blanco-Pillado:2021ygr}:
mono (orange), kink (green), cusp (blue), and a spectrum computed from a
simulated gravitational backreaction model (magenta).
}
{Lower left: Contour representing the $n_T-r$ (tensor spectral index
and tensor-to-scalar ratio, respectively) parameter space consistent with the
NANOGrav 12.5-year 5-frequency power-law $2\sigma$ confidence contour considering the {\it Bicep2-Keck Array} and {\it Planck} constraint that $r < 0.07$ \cite{Vagnozzi:2020gtf}.
} 
{Lower right: NANOGrav 12.5-year contours for SMBHBs, which are expected
to have a spectral index of $\gamma_{\rm CP} = 13/3$, corresponding to
$A_{\rm CP}=[1.4, 2.7] \times 10^{-15}$ from the NANOGrav 12.5-year 5-frequency
and broken power law $2\sigma$ contours.
}
}
\label{fig:SourceContours}
\end{figure*}

Finally, we compare the results for two types of purely hydrodynamic
turbulence with vortical and irrotational forcings of \Tab{Tsummary_hydro}.
The result is shown in \Fig{pspecm_other}.
All these cases are for plane wave forcings.
For irrotational forcing, we do not see the sharp drop-off of spectral
power for frequencies above the peak value as in the vortical case. 
This suggests that in the inertial range of irrotational turbulence,
there is still some power to contribute to GW driving compared with the
vortical case, where this is almost not possible at all. 
However, the spectrum in the irrotational case shows a fairly steep
spectrum proportional to $f^{-7}$, so the effect on GW production is
here also rather weak. 
Nevertheless, the spectral form of the peak might give interesting
diagnostic clues about the nature of turbulent driving at the time of
GW production.

We mention in passing that in earlier work, it was found that
irrotational turbulence is much more efficient in driving GWs
than vortical turbulence \cite{Pol:2019yex}.
Remarkably, here this is no longer the case and vortical and
irrotational turbulence have rather similar GW energies.
This could be related to the small value of $\kf$, possibly
combined with a comparatively short time of driving.
However, to clarify this further, more targeted numerical
experiments would need to be performed.

To put our results into perspective, we compare in
Fig.~\ref{fig:SourceContours} with the contours for possible sources
of GWs in the nHz range in terms of $h_0^2\OmGW$ and $f$.
In the upper left, we show limits on GWs from magnetohydrodynamic (MHD) turbulence at QCD for
$\kf$ from 2 to 6 (corresponding to Runs~noh5,6, hel5,6, and noh7,8)
around the NANOGrav sensitivity range.
Additionally, we show the {\it integrated} bounds on $h_0^2\OmGW$
from the CMB and BBN \cite{MaggioreText2}, noting that the actual bound
on the peak of $h_0^2\OmGW$ can fall above these lines.
Contours in the upper right correspond to four different models for
the average power spectrum of GWs from a network of cosmic strings
of different tensions \cite{Blanco-Pillado:2021ygr}.
The bottom left contour corresponds to the parameter space of $n_T-r$
(tensor spectral index and tensor-to-scalar ratio) consistent with the
$2\sigma$ contours of the NANOGrav 12.5-year 5-frequency power law.
This corresponds to Figs.~1 and 2 of Ref.~\cite{Vagnozzi:2020gtf}.
The lower right contour represents the NANOGrav 12.5-year 2$\sigma$
contours for the 5-frequency and broken power laws for a population of
SMBHBs, expected to have a spectral index of $\gamma_{\rm CP} = 13/3$.
In this work we found $\OmGW(f)\sim f^{1/2}$ for nonhelical turbulence,
corresponding to a spectral index of $\gamma_{\rm CP} = 4.5$, which
falls at the edge of the $1\sigma$ confidence contours for the NANOGrav
5-frequency and broken power laws.
The scaling $\OmGW(f)\sim f$, which we found in the case of helical 
turbulence, corresponds to $\gamma_{\rm CP} = 4$ and falls within
the $2\sigma$ contours from NANOGrav.
The spectral index for SMBHBs, $\gamma_{\rm CP} = 13/3$ falls between
these.

\section{Conclusions}
\label{Concl}

In the present work, we have shown that the magnetic stress from hydrodynamic and MHD
turbulence with scales comparable to the cosmological horizon scale at the
time of the QCD phase transition can drive GWs in the range accessible
to NANOGrav, if the magnetic energy density is 3--10\% of the radiation
energy density.
The low-frequency tail below the peak frequency at $10\nHz$ or so is
shallower in the nonhelical case than in the helical one, i.e.,
$\propto f^{1/2}$ compared to $\propto f$.
Both scalings are, however, shallower than what was expected based on
earlier analytical calculations.
Also the inertial range spectrum above the peak is shallower without
helicity than with, but here, both spectra are steeper than what
is expected if the GW spectrum was a direct consequence of the MHD 
turbulence spectrum \cite{Pol:2019yex,BB20}.
The reason for this is primarily the relatively short time of turbulent 
driving (one Hubble time).
It is short compared with the turnover time which,
for our runs with the smallest $\kf$ of two, is much longer:
16 (100) Hubble times for our runs without (with) helicity.
Therefore, there was not enough time to fully establish the GW spectrum
at high wave numbers. 
For our earlier runs with larger values of $\kf$, this effect was less
pronounced than for smaller values of $\kf$, but it is still quite noticeable,
especially in the helical case where forward cascading is weaker than
in the nonhelical case. 

Our work has led to new insights regarding the possibility of using
an observed GW spectrum for making statements about the nature of the
underlying turbulence in the early universe.
One is the already mentioned slope of the subinertial range spectrum.
Another is the position of the peak of the spectrum.
Finally, there is the strength of the drop of the spectral power for
frequencies above the peak frequency, and the subsequent slope after
the drop, which is most likely too small to be detectable.
This, however, depends on the duration of turbulent driving and could
be higher if the driving time was longer. 
The specific features of the spectrum near the peak are different
for helical and nonhelical turbulence.
This could, in principle, give information about the presence 
of parity violation, when would also lead to circularly polarized
GWs. 

\vspace{2mm}
Data availability---The source code used for the
simulations of this study, the {\sc Pencil Code},
is freely available from Ref.~\cite{PC}.
The simulation setups and the corresponding data
are freely available from Ref.~\cite{DATA}.

\acknowledgements
Support through the Swedish Research Council, grant 2019-04234,
and Shota Rustaveli GNSF  (grant FR/19-8306)
are gratefully acknowledged.
We acknowledge the allocation of computing resources provided by the
Swedish National Allocations Committee at the Center for Parallel
Computers at the Royal Institute of Technology in Stockholm.


\end{document}